\documentclass[prd,showpacs,preprintnumbers,amsmath,amssymb,superscriptaddress,floatfix,nofootinbib]{revtex4}
\usepackage{graphicx}
\usepackage{epstopdf}
\usepackage{amsmath}
\usepackage{amsfonts}
\usepackage{amssymb}
\usepackage{bm}

\begin{document}

\title{$Z_{cs}$ states from the $D^{*}_s \bar D^*$ and $J/\psi K^*$ coupled channels: Signal in $B^+ \to J/\psi \phi K^+$ decay}

\author{Natsumi Ikeno}
\email{ikeno@tottori-u.ac.jp}
\affiliation{Department of Agricultural, Life and Environmental Sciences, Tottori University, Tottori 680-8551, Japan}

\author{Raquel Molina}
\email{raquel.molina@ific.uv.es}
\affiliation{Departamento de F\'{\i}sica Te\'orica and IFIC,
Centro Mixto Universidad de Valencia-CSIC Institutos de Investigaci\'on de Paterna, Aptdo.22085, 46071 Valencia, Spain}

\author{Eulogio Oset}
\email{oset@ific.uv.es}
\affiliation{Departamento de F\'{\i}sica Te\'orica and IFIC,
Centro Mixto Universidad de Valencia-CSIC Institutos de Investigaci\'on de Paterna, Aptdo.22085, 46071 Valencia, Spain}
\date{\today}

\begin{abstract}
We study the $D^{*}_s \bar D^*$ system in connection with the $J/\psi K^*$ in coupled channels and observe that, within reasonable values of the cut-off used to regularize the loops, the system does not develop a bound state. However, the $J^P = 2^+$ channel has enough attraction to create a strong cusp structure that shows up in the $J/\psi K^+$ invariant mass distribution in the $B^+ \to J/\psi \phi K^+$ decay at the $D^*_s \bar D^*$ threshold. Such structure is visible in the experimental $B^+ \to J/\psi \phi K^+$ and $\bar B^0 \to J/\psi K^+ K^-$ decays, with small statistics, and our results should stimulate further measurements around this region, given the fact that cusp effects provide as valuable information on hadron dynamics as resonances themselves.
\end{abstract}

\maketitle
\section{Introduction}
  The discovery of the $Z_{cs}(3985)$ state by the BESIII collaboration~\cite{BESIIIexp}
in the mass distribution of $\bar D^*_s D$ and $\bar D_s D^*$ added a new type of exotic meson, with $ \bar c s c \bar u$ to the increasing long list of exotic hadronic states (see reviews \cite{hosaka,nora,chenzhu,olsen,lebed,tomaz,ulfmole}). The reaction of the theoretical community has been fast, and many papers have been devoted to understanding the nature of this state. As usual, three lines have been followed to describe the state, assuming it to be a tetraquak state, a meson-meson molecular state, or using QCD sum rules. Independent on the picture used, it is an  extended idea that the $Z_{cs}(3985)$ state is an SU(3) partner of the $Z_c(3900)$ state, replacing a $q$ quark by a strange $s$ quark. In Refs.~\cite{Wan:2020oxt,Wang:2020rcx,Wang:2020iqt,Azizi:2020zyq,Xu:2020evn,Wang:2020dgr,Albuquerque:2021tqd,Ozdem:2021yvo,Chen:2021erj} the QCD sum rules method is used to generate the $Z_{cs}$ and the state is obtained as a $\bar D^*_s D$ configuration, as a strangeness analog of the $\bar D^* D$ configuration of the $Z_{c}$, with the usual large uncertainties about the mass of the states. Tetraquark calculations, both using diquark-antidiquark configurations 
$[sc][\bar q \bar c]$ \cite{Chen:2021uou,Shi:2021jyr,Giron:2021sla} or diquark-antidiquark and meson-meson  configurations~\cite{Jin:2020yjn,david} have been carried out. In Ref.~\cite{Jin:2020yjn} the molecular component is shown not to bind and the diquark picture is favoured. In Ref.~\cite{david} the chiral constituent quark model is used and the molecular component, including coupled channels, is shown to lead to a virtual state. A different analysis along these lines is carried out in Ref.~\cite{rosner} where the $Z_{cs}(3985)$ of Ref.~\cite{BESIIIexp} and the $Z_{cs}(4000)$  observed by the LHCb collaboration in Ref.~\cite{LHCb:2021uow} are supposed to be two different states and follow a mixing like the one that mixes the $K_1(1270)$ and $K_1(1440)$ states.\footnote{This picture would actually be more complicated if one considers the existence of two $K_1(1270)$ states coupling differently to $K^* \pi$ and $\rho K$,
and appearing at different energies, as found in the chiral unitary approach in Refs.~\cite{roca,rocageng}.}. 

  Much work is done considering molecules using directly dynamics for meson meson interaction~\cite{Meng:2020ihj,Wang:2020kej,Yang:2020nrt,Chen:2020yvq,Cao:2020cfx,Du:2020vwb,Sun:2020hjw,Wang:2020htx,ikeno,Yan:2021tcp,Meng:2021rdg,Ding:2021igr,Wu:2021ezz,Hidalgo-Duque:2012rqv,Baru:2021ddn}. By analogy to the $Z_c(3900)$, the SU(3) partner in the strange sector $Z_{cs}(3985)$ is favored as the $D^{*-}_s D^0+D^-_s D^{*0}$ combination. One exception is Ref.~\cite{Meng:2021rdg}, where this combination is preferred for the $Z_{cs}(4000)$ state, while the $D^{*-}_s D^0 - D^-_s D^{*0}$ combination is proposed for the $Z_{cs}(3985)$, all that assuming that the $Z_{cs}(3985)$, $Z_{cs}(4000)$ are different states, something not supported in Ref.~\cite{david} where the two states are claimed to be the same one. 

  Even admitting the same molecular picture, different works use different dynamics. In Ref.~\cite{Chen:2020yvq} $\pi$ and $\eta$ exchange are considered and the interaction is found too weak to bind. In Ref.~\cite{Sun:2020hjw} the local hidden gauge approach is used and a pole is found in the third Riemann sheet, rather than the ordinary second sheet, indicating not much binding. A clarification of the issue is provided in Refs.~\cite{ikeno,Dong:2021juy}, where heavy quark symmetry is assumed and the source of the interaction is the exchange of vector mesons. The works follow the basic line of Ref.~\cite{juanxiao} where the heavy quark spin symmetry is assumed, which implies relationships between the different transition potentials and the dynamics is taken from the local hidden gauge approach~\cite{hidden1,hidden2,hidden4,hideko}. The exchange of light vectors is shown to respect heavy quark spin symmetry because the heavy quarks in the mesons act as spectators \cite{liangxiao}. Both in the case of the $Z_c(3900)$ \cite{aceti} and in the case of the  $Z_{cs}(3985)$, the diagonal interaction with the exchange of light vectors is zero. In Ref.~\cite{Dong:2021juy} only single channels are considered, and lacking the dominant terms from the exchange of light vectors, only a virtual state has some room around threshold.  The presence of a threshold together with some attractive interaction, even if weak, can lead to some structure around threshold as discussed in Ref.~\cite{Dong:2020hxe}. The interaction including coupled channels leads to a stronger attraction than in single channels, and as shown in Ref.~\cite{ikeno} is strong enough in the $D^{*-}_s D^0+D^-_s D^{*0}$ combination to produce a mass distribution for $D^{*-}_s D^0$ and $D^-_s D^{*0}$ compatible with experiment, while the $D^{*-}_s D^0-D^-_s D^{*0}$ combination is unable to reproduce the experimental shape. Due to the weak primary interaction, in Ref.~\cite{Yan:2021tcp}  the exchange of the $a_1(1260)$ is evaluated providing a small contribution that helps in the binding. A different kind of approach is used in Ref.~\cite{Guo:2020vmu} using an effective range expansion to justify that the $Z_{cs}(3985)$ is the SU(3) partner of the  $Z_c(3900)$. Also in Ref.~\cite{Ge:2021sdq} a discussion is conducted suggesting that the signal of the $Z_{cs}(3985)$ could be due to threshold cusps enhanced by a possible triangle singularity. 

     In the present work we extend the molecular picture to the $D^{*-}_s D^{*0}$,  $D^{*-}_s D^{*0}$ and coupled channels sector as shown in detail in Ref.~\cite{Hidalgo-Duque:2012rqv}. Within the heavy quark spin symmetry assumptions there is a trivial mapping of the interaction in the two sectors. However, in the charm sector nonleading terms or the interaction in the large $N_c$ counting are not negligible, and as we shall see, in the absence of leading diagonal terms coming from the exchange of light vectors, the contact terms and the exchange of heavy $D^*_s$ vectors provide a sizable interaction, which is strong enough to produce some bound states, or threshold structures, also removing the degeneracy of the $0^+,1^+,2^+$ states. 

We construct the $J/\psi K$ mass distributions in the $B^+ \to J/\psi \phi K^+$ decay and show that a possible narrow peak observed in Ref.~\cite{LHCb:2021uow} (see the region around 4120~MeV between the $Z_{cs}(4000)$ and $Z_{cs}(4220)$ peaks in the $J/\psi K^+$ mass distribution of Fig.~3) 
can be naturally associated to a near bound (virtual) $D^*_s \bar D^*$ state with $J^{PC} = 2^{++}$.

\section{Formalism for the interaction}\label{sec:form}
For the $D^*_s \bar D^*$ we consider as coupled channels, 1, 2, 
\begin{equation}
 D_s^{*+} \bar D^{*0}~~(1), ~~~~~ J/\psi K^{*+}~~(2), 
\label{eq:channel}
\end{equation}
and take the interaction from the extension of the local hidden gauge approach~\cite{hidden1,hidden2,hidden4,hideko} to the charm sector. The contact term  is given by
\begin{equation}
{\cal L}^{(c)}=\frac{g^2}{2}\langle V_\mu V_\nu V^\mu V^\nu-V_\nu V_\mu
V^\mu V^\nu\rangle\ ,
\end{equation}
with $g= M_V/{2f}$ ($M_V=800$~MeV, $f=93$~MeV) with $\langle \rangle$ indicating the trace of the matrices and
\begin{equation}
\renewcommand{\tabcolsep}{1cm}
V_\mu=\left(
\begin{array}{cccc}
\frac{\rho^0}{\sqrt{2}}+\frac{\omega}{\sqrt 2} & \rho^+ & K^{*+}&\bar{D}^{*0}\\
\rho^- & -\frac{\rho^0}{\sqrt{2}}+\frac{\omega}{\sqrt 2} & K^{*0}&D^{*-}\\
K^{*-} & \bar{K}^{*0} &\phi&D^{*-}_s\\
D^{*0}&D^{*+}&D^{*+}_s&J/\psi\\
\end{array}
\right)_\mu.
\label{eq:vmu}
\end{equation}
There is another source of the interaction given by the exchange of vectors based on the three vector vertex
\begin{equation}
 {\cal L}_{VVV}=ig\langle (V^\mu \partial_\nu V_\mu -\partial_\nu V_\mu V^\mu)
V^\nu)\rangle.
\label{eq:L_VVV}
\end{equation}

Working close to threshold allows one to take the approximation of neglecting the three momenta of the external vectors, which implies $\epsilon^0 = 0$ for the external vectors. This implies that $V^\nu$ in Eq.~(\ref{eq:L_VVV}) cannot be external since $\nu = 1,2,3$ will involve three vectors through $\partial_\nu$. Thus, the field $V^\nu$ corresponds to the exchanged vector and the interaction vertex is formally like the $VPP$ ($P$ for pseudoscalars) with the additional factor $(-) \epsilon_\mu \epsilon^{'\mu} = \vec{\epsilon}\cdot \vec{\epsilon}\,^{'}$ for the external vectors.
We have to evaluate the contact term of the diagrams.
 
\begin{figure}[h]
 \centering
 \includegraphics[scale=0.7]{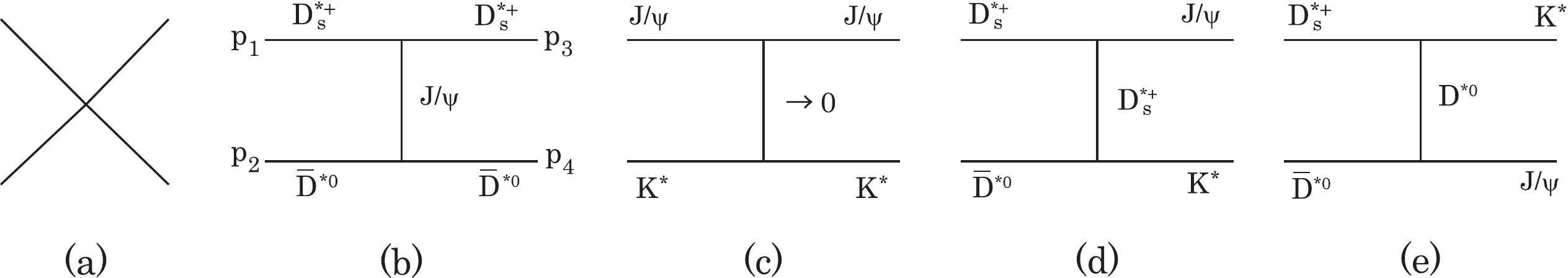}
 \caption{
Contact term (a) and vector exchange terms (b, c, d, e) involved in the interaction of the coupled channels.}
 \label{fig:fig1}
\end{figure}

We work with the interaction in $s$-wave, but due to the spins of the vertex we can have now spins $J=0,1,2$ and we must separate the interaction for each of this total spin sectors. This is done with the projectors ${\cal P}^{(0)}, {\cal P}^{(1)}, {\cal P}^{(2)}$ of Refs.~\cite{raquelrho,raquelxyz},
\begin{eqnarray}
&&{\cal P}^{(0)}= \frac{1}{3}\epsilon_\mu \epsilon^\mu \epsilon_\nu \epsilon^\nu\nonumber\\
&&{\cal P}^{(1)}=\frac{1}{2}(\epsilon_\mu\epsilon_\nu\epsilon^\mu\epsilon^\nu-\epsilon_\mu\epsilon_\nu\epsilon^\nu\epsilon^\mu)\nonumber\\
&&{\cal P}^{(2)}=  \frac{1}{2}(\epsilon_\mu\epsilon_\nu\epsilon^\mu\epsilon^\nu+\epsilon_\mu\epsilon_\nu\epsilon^\nu\epsilon^\mu)-\frac{1}{3}\epsilon_\mu\epsilon^\mu\epsilon_\nu\epsilon^\nu .
\label{eq:projmu}
\end{eqnarray}

Evaluating the contact term for the interaction of the coupled channels of Eq.~(\ref{eq:channel}), we obtain
\begin{eqnarray}
V^{(c)} &=& g^2\left(
\begin{array}{cc}
2 & -4 \\
-4 & 0 \\
\end{array}
\right), ~~~~J=0, \label{eq:Vc_J0}\\
V^{(c)} &=& g^2\left(
\begin{array}{cc}
3 & 0 \\
0 & 0 \\
\end{array}
\right), ~~~~J=1, \label{eq:Vc_J1} \\
V^{(c)} &=& g^2\left(
\begin{array}{cc}
-1 & 2 \\
2 & 0 \\
\end{array}
\right), ~~~~J=2. 
\end{eqnarray}
We can see that the $V^{(c)}$ interaction is different for the different spin channels. The diagonal terms are repulsive or zero for $J=0, 1$, but attractive for $J=2$. This gives us chances that one can find some binding in $J=2$, which is usually the most bound channel in all cases of the $VV$ interaction~\cite{raquelrho,raquelxyz,gengvec}.

From the exchange of vectors in Fig.~\ref{fig:fig1} we obtain the result
\begin{eqnarray}
V^{\rm (ex)} &=& g^2\left(
\begin{array}{cc}
-\frac{(p_1 + p_3)\cdot (p_2 + p_4)}{M^2_{J/\psi}} & -\frac{3s- \sum M^{'2}_{i}}{D} \\
-\frac{3s-\Sigma M^{'2}_{i}}{D} & 0 \\
\end{array}
\right), ~~~~J=0,2 \\
V^{\rm (ex)} &=& g^2\left(
\begin{array}{cc}
-\frac{(p_1 + p_3)\cdot (p_2 + p_4)}{M^2_{J/\psi}} & 0 \\
0 & 0 \\
\end{array}
\right), ~~~~J=1, 
\label{eq:Vex_J1}
\end{eqnarray}
where 
\begin{equation}
 \Sigma M^{'2}_{i} = M^2_{D^{*}_s} +  M^2_{\bar D^{*0}} +  M^2_{J/\psi} +  M^2_{K^*}, ~~~ D = M^2_{J/\Psi}- 2 \bar{M}_{D^{*}} E_{J/\psi}
\end{equation}
with 
\begin{equation}
\bar{M}_{D^{*}} = \frac{1}{2} (M_{D^{*}_s} + M_{\bar D^{*}} ),~~~ E_{J/\psi} = \frac{s + M^2_{J/\psi} -M^2_{K^*} }{2 \sqrt{s}} ,
\end{equation}
where $p_i$ refer to the momenta of the particles as shown in Fig.~\ref{fig:fig1} and to simplify the formulas we have taken an average $M_{\bar D^*}$ between $D^*_s$ and $D^{*0}$ in the denominator $D$. The product $(p_1 + p_3) \cdot (p_2 + p_4)$ must be projected in $s$-wave, with the results~\cite{roca}
\begin{eqnarray}
 (p_1 + p_3) \cdot (p_2 + p_4) \to \frac{1}{2} \left\{
3s - \sum_i M^2_{i} - \frac{1}{s} (M^2_1 - M^2_2) (M^2_3 - M^2_4)
\right\},
\end{eqnarray}
where now  $M_i (i=1,2,3,4)$ refer to the particles in the diagrams of Fig.~\ref{fig:fig1} with the order expressed there.

\section{Decay channels}
\begin{figure}[tbh]
 \centering
 \includegraphics[scale=0.8]{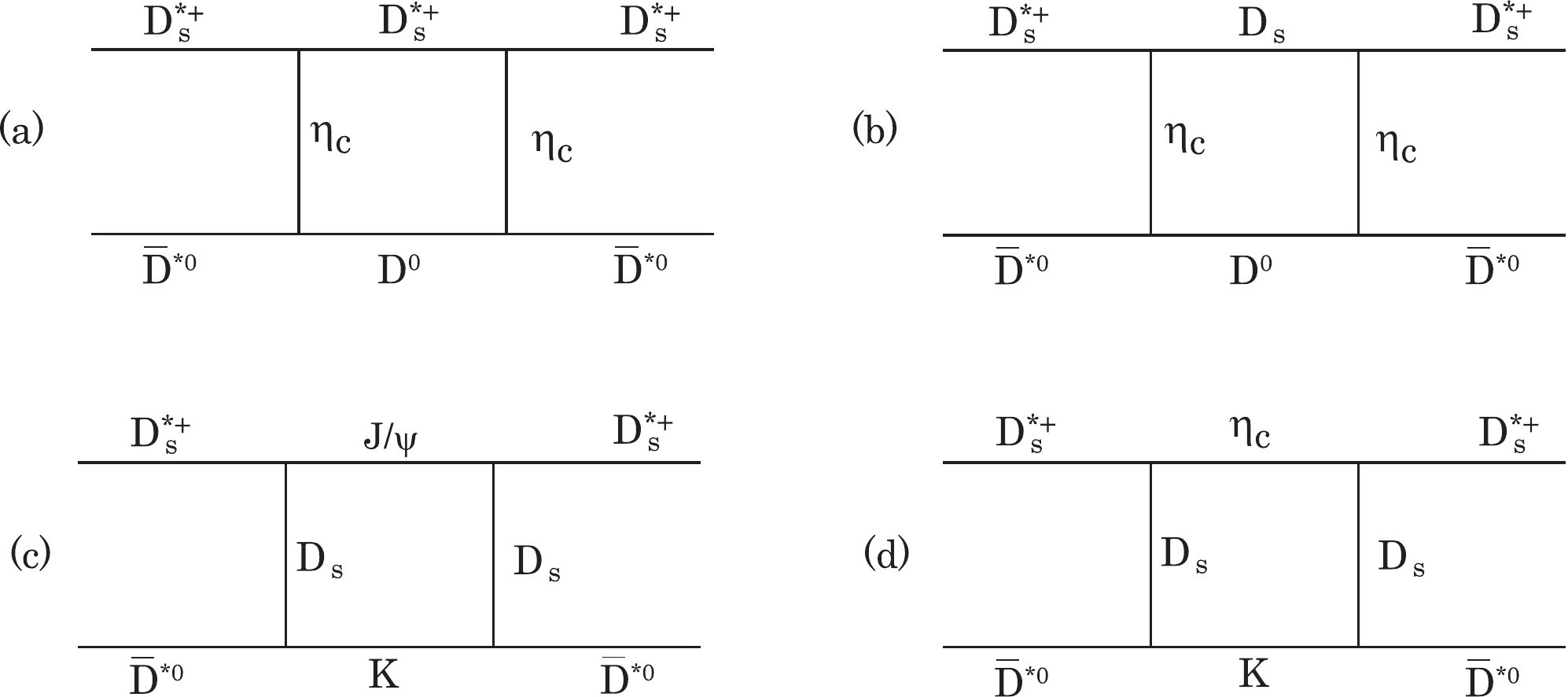}
 \caption{Diagrams containing the decay channels in the intermediate states. The threshold of the intermediate states are (a) 3976~MeV; (b) 3832~MeV; (c) 3593~MeV; (d) 3477~MeV. }
\label{fig:fig2}
\end{figure}

We should note that the $J/\psi K^*$ state is 129~MeV below the $D^*_s \bar D^{*0}$ threshold, hence any state that we find around the $D^*_s \bar D^{*0}$ threshold will decay to $J/\psi K^*$, except for $J=1$ where the transition potential is zero. The case is irrelevant since the interaction is repulsive in that channel. Let us study the decay in other channels. We only study the decay channels of the $D_s^{*} \bar D^{*0}$ component, which is the relevant one in the states that we obtain. We can look at the decay channels of Fig.~\ref{fig:fig2} which are not of $VV$ type.

It is clear that given the higher thresholds of the diagrams (a), (b) of Fig.~\ref{fig:fig2} and the ratio of propagator $(M_{D_s}/M_{\eta_c})^2 = 0.19$, the relevant decay channels correspond to diagrams (c), (d), which are the only ones that we consider. Taking into account angular momentum and parity conservation we have the results of Table~\ref{table:tab1}.

\begin{table}[t]
 \begin{center}
\caption{Possible values of $L$ for $J/\psi K$ and $\eta_c K$ which make the transition of $D^{*+}_s \bar D^{*0}$ to $J/\psi K$ or $\eta_c K$ possible. The  $\times$ symbol indicates that the transition is forbidden.}
  \begin{tabular}{l|l|l}
  \hline
   $J^P$~~ &~~~~~ $J/\psi K$ &~~~~~ $\eta_c K$\\
   \hline
   $2^+$ &~~~ $1^-, 0^-$; $L=2$  &~~~ $0^-, 0^-$; $L=2$ \\ 
   $1^+$ &~~~ $1^-, 0^-$; $L=0$  &~~~ $0^-, 0^-$; $L=1$  $\times$ \\
   $0^+$ &~~~ $1^-, 0^-$; $L=1$ $\times$  &~~~ $0^-, 0^-$; $L=0$ \\
   \hline
  \end{tabular}
\label{table:tab1}
 \end{center}
\end{table}

We can see that for $J^P = 2^+$ both $J/\psi K$ and $\eta_c K$ decay channels are possible. For $1^+$ only the decay to $J/\psi K$ is possible and for $0^+$ only the decay to $\eta_c K$ is possible.

\subsection{$\eta_c K$ intermediate state}
\begin{figure}[!h]
 \centering
 \includegraphics[scale=0.85]{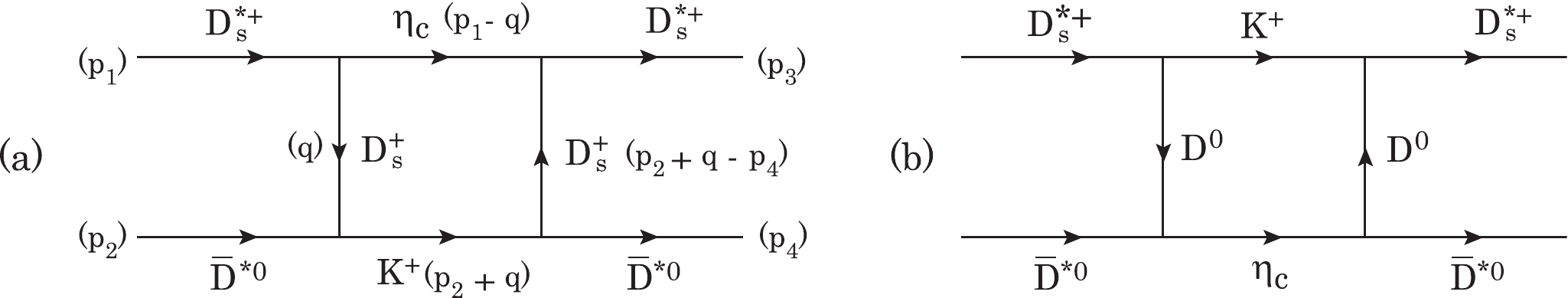}
 \caption{Diagram for $D^{*+}_s \bar D^{*0}$ decay to $\eta_c K^+$.}
\label{fig:fig3}
\end{figure}

We shall evaluate the contribution of the diagram of Fig.~\ref{fig:fig3} the $D^{*+}_s \bar D^{*0} \to D^{*+}_s \bar D^{*0}$ potential.

The relevant vertices are $VPP$ given by the Lagrangian
\begin{equation}
{\cal L}_{VPP}=-ig\langle [ P,\partial_\mu P]V^\mu\rangle  
\end{equation}
with $P$ given now by the matrix~\cite{ikeno},
\begin{equation}
P=\left(
\begin{array}{cccc}
\frac{\eta}{\sqrt{3}}+\frac{\eta'}{\sqrt{6}}+\frac{\pi^0}{\sqrt{2}} & \pi^+ & K^+&\bar{D}^0\\
\pi^- &\frac{\eta}{\sqrt{3}}+\frac{\eta'}{\sqrt{6}}-\frac{\pi^0}{\sqrt{2}} & K^{0}&D^-\\
K^{-} & \bar{K}^{0} &-\frac{\eta}{\sqrt{3}}+\sqrt{\frac{2}{3}}\eta'&D^-_s\\
D^0&D^+&D^+_s&\eta_c
\end{array}
\right)\ .
\label{eq:pfields}
\end{equation}
By neglecting the $\epsilon^0$ components of the vector, as done before, we obtain for the two diagrams of Fig.~\ref{fig:fig3}.
\begin{eqnarray}
 -i t^{(a)} &=&  \int \frac{d^4 q}{(2\pi)^4}  \frac{i}{q^2 - M^2_{D_s}+ i\epsilon}
\ 4 g^2 \ \epsilon_i(D_s^{*+}) \epsilon_j(\bar D^{*0})  \  q_i q_j
 \nonumber\\
& \cdot & 
 \frac{i}{(p_2 + q  -p_4)^2 - M^2_{D_s}+ i\epsilon}  
\ 4 g^2 \epsilon'_{\ell}(D^{*+}_s) \epsilon'_m(\bar D^{*0}) q_{\ell} q_{m}
\nonumber\\
& \cdot & 
\frac{i}{(p_1 - q)^2 - M^2_{\eta_c}+ i\epsilon} \
\frac{i}{(p_2 + q)^2  - M^2_{K}+ i\epsilon}.
\label{eq:t_a}
\end{eqnarray}
From experience~\cite{raquelrho,gengvec,toledo}, the real part of these box diagrams is small compared to the more important terms stemming from vector exchange, and hence we keep only the imaginary part of the diagrams that contains the new decay channels. Due to this, in the decomposition of the propagator into positive and negative energy parts ($\omega(q) = \sqrt{m^2 + \vec{q}~^2}$)
\begin{equation}
\frac{1}{q^2 - m^2} \equiv \frac{1}{2\omega(q)} \left( \frac{1}{q^0 - \omega(q) + i\epsilon} 
- \frac{1}{q^0 + \omega(q) - i\epsilon} \right) ,
\end{equation}
we take only the positive energy part for the intermediate $\eta_c$ and $K$ states, and because the $D_s, D$ states are massive we can equally neglect the negative energy part. This simplifies the expression of Eq.~(\ref{eq:t_a}) and we find
\begin{eqnarray}
 -i t^{(a)} &=& i \int \frac{d^4 q}{(2\pi)^4} (4 g^2)^2 \ \epsilon_i(D_s^{*+}) \epsilon_j(\bar D^{*0}) \epsilon'_{\ell}(D^{*+}_s) \epsilon'_m(\bar D^{*0}) \nonumber\\
& \cdot & q_i q_j q_{\ell} q_{m} \ 
\frac{1}{2 \omega_{D_s}(q)} \frac{1}{q^0 - \omega_{D_s}+ i\epsilon} \
\frac{1}{2 \omega_{D_s}(q)} \frac{1}{p^0_2 + q^0  -p^0_4 - \omega_{D_s}+ i\epsilon}  \nonumber\\
& \cdot & 
\frac{1}{2 \omega_{\eta_c}(q)} \frac{1}{p^0_1 - q^0 - \omega_{\eta_c}+ i\epsilon} \
\frac{1}{2 \omega_{K}(q)} \frac{1}{p^0_2 + q^0  - \omega_{K}+ i\epsilon},
\end{eqnarray}
with $\omega_i = \sqrt{m_i^2 + \vec{q}~^2 }$,
where the $q^0$ integration is immediately performed using Cauchy's residues, and one gets
\begin{eqnarray}
 t^{(a)} &=&  \int \frac{d^3 q}{(2\pi)^3} \frac{1}{2 \omega_{D_s}(q)} \frac{1}{2 \omega_{D_s}(q)}
\frac{1}{2 \omega_{\eta_c}(q)} \frac{1}{2 \omega_{K}(q)} \nonumber\\
& \cdot & 
(4 g^2)^2 \ \epsilon_i(D_s^{*+}) \epsilon_j(\bar D^{*0}) \epsilon'_{\ell}(D^{*+}_s) \epsilon'_m(\bar D^{*0}) 
\  q_i q_j q_{\ell} q_{m} \ 
\nonumber\\
& \cdot & 
\left( \frac{1}{p^0_1 - \omega_{\eta_c}(q) -\omega_{D_s}(q) + i\epsilon }\right)^2
\frac{1}{\sqrt{s} - \omega_{\eta_c}(q) - \omega_K(q) + i\epsilon}.
\end{eqnarray}

A further simplification can be done using
\begin{equation}
  \int \frac{d^3q}{(2\pi)^3} f(\vec{q}\,^2) q_i q_j q_\ell q_m 
= \int \frac{d^3q}{(2\pi)^3} f(\vec{q}\,^2) \vec{q}\,^4 \frac{1}{15} (\delta_{ij} \delta_{\ell m} + \delta_{i \ell}\delta_{jm} + \delta_{im} \delta_{j\ell})\ ,
\end{equation}
which leads to the combination of polarization vectors in order $1,2,3,4$ of the ordering of Fig.~\ref{fig:fig3}
\begin{equation}
 \epsilon_i \epsilon_i  \epsilon_j \epsilon_j + \epsilon_i \epsilon_j \epsilon_i \epsilon_j + \epsilon
_i \epsilon_j \epsilon_j \epsilon_i, 
\end{equation}
which by virtue of the spin projectors of Eq.~(\ref{eq:projmu}) gives
\begin{equation}
 3 {\cal P}^{(0)} + {\cal P}^{(0)} + {\cal P}^{(1)} + {\cal P}^{(2)} + {\cal P}^{(2)} - {\cal P}^{(1)} + {\cal P}^{(0)} 
= 5 {\cal P}^{(0)} + 2 {\cal P}^{(2)} ,
\end{equation}
and ${\cal P}^{(1)}$ does not appear as anticipated in Table~\ref{table:tab1}.
Finally we obtain an easy formula for $t^{(a)}$
\begin{eqnarray}
 t^{(a)} &=&  W_s \frac{1}{15} (4 g^2)^2 \
\int \frac{d^3 q}{(2\pi)^3} \left( \frac{1}{2 \omega_{D_s}(q)} \right)^2
\frac{1}{2 \omega_{\eta_c}(q)} \frac{1}{2 \omega_{K}(q)} \ \vec{q}\,^4 \nonumber\\
& \cdot & 
\left( \frac{1}{p^0_1 - \omega_{\eta_c}(q) -\omega_{D_s}(q) + i\epsilon }\right)^2
\frac{1}{\sqrt{s} - \omega_{\eta_c}(q) - \omega_K(q) + i\epsilon},
\label{eq:t_a2}
\end{eqnarray}
where 
\begin{equation}
\renewcommand{\tabcolsep}{1cm}
W_s =\left\{ 
\begin{array}{c}
5\\
0\\
2\\
\end{array}
\right\}
\begin{array}{c}
~~~J=0\\
~~~J=1\\
~~~J=2\\
\end{array}
\end{equation}
The imaginary part of Eq.~(\ref{eq:t_a2}) is readily obtained and we find:
\begin{eqnarray}
 {\rm Im} V^{(a)} &=&  - W_s \frac{1}{15} (4 g^2)^2 \frac{1}{8\pi} \frac{1}{\sqrt{s}}\, q^5 
\left( \frac{1}{2 \omega_{D_s}(q)} \ \frac{1}{p^0_1 - \omega_{\eta_c}(q) -\omega_{D_s}(q) }\right)^2
F(q)^4 \,  F_{\rm HQ}  ,
\end{eqnarray}
where 
\begin{equation}
 q = \frac{\lambda^{1/2}(s, M^2_{\eta_c}, M^2_K)}{2 \sqrt{s}};~~~~
 p^0_1 = \frac{s + M^2_{D^*_s} - M^2_{\bar D^{*0}}}{2 \sqrt{s} }   ,
\end{equation}
and we have added a form factor for each vertex as in Ref.~\cite{molinaoset},
\begin{equation}
F(q)= e^{q^2/\Lambda^2} =e^{(q^{02} - \vec{q}\,^{2})/\Lambda^2}\ ,
~~~ q^0 = p^0_1 - \omega_{\eta_c}(q),
\end{equation}
with $\Lambda = 1200$~MeV, and the factor $F_{\rm HQ}$ required by the normalization we use with the heavy quarks and corrections needed in the space components of the $VPP$ vertices that we use \cite{liangxiao} ,
\begin{equation}
 F_{\rm HQ} = \left( \frac{M_{D^*}}{M_{K^*}} \right)^4.
\end{equation}

Following the same steps we find 
\begin{eqnarray}
 {\rm Im} V^{(b)} &=&  - W_s \frac{1}{15} (4 g^2)^2 \frac{1}{8\pi} \frac{1}{\sqrt{s}}\, q'\,^5 
\left( \frac{1}{2 \omega_{D^0}(q')} \ \frac{1}{p^0_1 - \omega_{K}(q') -\omega_{D^0}(q') }\right)^2
F(q')^4 \,  F_{\rm HQ}  ,
\end{eqnarray}
where 
\begin{equation}
F(q')= e^{q'\,^2/\Lambda^2} =e^{(q'\,^{02} - \vec{q'}\,^{2})/\Lambda^2}\ ,
~~~ q'\,^0 = p^0_1 - \omega_{K}(q'),
\end{equation}
\begin{equation}
 q' = \frac{\lambda^{1/2}(s, M^2_{\eta_c}, M^2_K)}{2 \sqrt{s}} = q. 
\end{equation}
The two expressions for Im$V^{(a)}$, Im$V^{(b)}$ are now added to the diagonal $D^{*+}_s \bar D^{*0}$, $D^{*+}_s \bar D^{*0}$ as 
\begin{equation}
V_{D^{*+}_s \bar D^{*0}, D^{*+}_s \bar D^{*0}} \rightarrow V_{D^{*+}_s \bar D^{*0}, D^{*+}_s \bar D^{*0}} + i {\rm Im}V^{(a)} + i {\rm Im}V^{(b)}
\end{equation}

\subsection{$J/\psi K$ intermediate state}
\begin{figure}[h]
 \centering
 \includegraphics[scale=0.85]{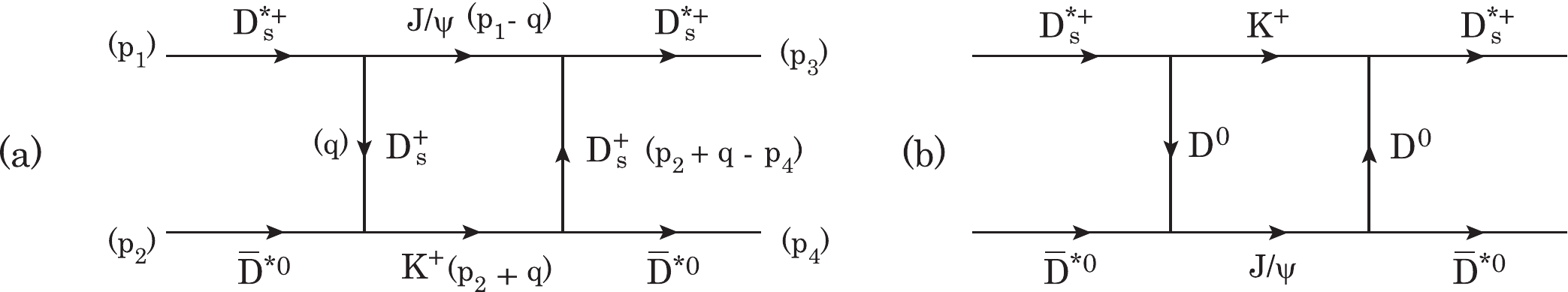}
 \caption{Diagram for $D^{*+}_s \bar D^{*0}$ decay to $J/\psi K$.}
\label{fig:fig4}
\end{figure}

The diagrams accounting for this intermediate state are shown in Fig.~\ref{fig:fig4}.
The vertices now involve the anomalous $VVP$ couplings. The Lagrangian is now given by Refs.~\cite{bramon,pelaez}.

\begin{equation}
 {\cal L}=\frac{G'}{\sqrt{2}}\epsilon^{\mu\nu\alpha\beta}\langle \partial_\mu V_\nu \partial_\alpha V_\beta P\rangle,
\end{equation}
with $G'=\frac{3g'}{4\pi^2f};g'=-\frac{G_Vm_\rho}{\sqrt{2}f^2}$, $G_V = 55$~MeV, $f=93$~MeV.

Following the same steps as in the former subsection we obtain now:
\begin{eqnarray}
 {\rm Im} V^{'(a)} &=&  - W'_s \, \frac{1}{8\pi} \frac{1}{\sqrt{s}} \left( \frac{G' g M_{D^*_s}}{\sqrt{2}} \right)^2  \frac{4}{15} \, q^5 
\left( \frac{1}{2 \omega_{D_s}(q)} \ \frac{1}{p^0_1 - \omega_{J/\psi}(q) -\omega_{D_s}(q)}\right)^2
F(q)^4 \,  F'_{\rm HQ}  .
\label{eq:ImVpa}
\end{eqnarray}
Now $F'_{\rm HQ} $ is different 
\begin{equation}
 F'_{\rm HQ} = \left( \frac{M_{D^*}}{M_{K^*}} \right)^2,
\end{equation}
because we have two anomalous couplings which are proportional to the external vector masses and do not require correction~\cite{liangxiao}.
In Eq.~(\ref{eq:ImVpa}), $W'_s$ is given by
\begin{equation}
\renewcommand{\tabcolsep}{1cm}
W'_s =\left\{ 
\begin{array}{c}
0\\
5\\
3\\
\end{array}
\right\}
\begin{array}{c}
~~~J=0\\
~~~J=1\\
~~~J=2,\\
\end{array}
\label{eq:Wps}
\end{equation}
\begin{equation}
q = \frac{\lambda^{1/2}(s, M^2_{J/\psi}, M^2_K)}{2 \sqrt{s}};~~~~
q^0 = p^0_1 - \omega_{J/\psi}(q)
\end{equation}
and
\begin{eqnarray}
 {\rm Im} V^{'(b)} &=&  - W'_s \, \frac{1}{8\pi} \frac{1}{\sqrt{s}} \left( \frac{G' g M_{D^*_s}}{\sqrt{2}} \right)^2  \frac{4}{15} \, q^5 
\left( \frac{1}{2 \omega_{D^0}(q)} \ \frac{1}{p^0_1 - \omega_{K}(q) -\omega_{D^0}(q)}\right)^2
F(q)^4 \,  F'_{\rm HQ}  ,
\end{eqnarray}
and finally we include all these decay channels taking for the $D_s^{*+} \bar D^{*0} \to D_s^{*+} \bar D^{*0}$ transition,
\begin{equation}
 V_{D_s^{*+} \bar D^{*0} \to D_s^{*+} \bar D^{*0}} + i \mathrm{Im} V^{(a)} + i \mathrm{Im} V^{(b)} + i \mathrm{Im} V^{'(a)} + i \mathrm{Im} V^{'(b)}.
\label{eq:V_all}
\end{equation}
Note that $W'_s = 0$ for $J=0$ in agreement with the findings of Table~\ref{table:tab1}.

\subsection{$J/\psi K$ distribution in $B^+ \to J/\psi \phi K^+$ decay}
In Ref.~\cite{LHCb:2021uow}, the $B^+ \to  J/\psi \phi K^+$ decay is studied and several mass distributions are shown. Two clear peaks are seen which are associated to the states $Z_{cs}(4000)$ and $Z_{cs}(4220)$. The $Z_{cs}(4000)$ state could correspond to the BESIII $Z_{cs}(3985)$, but the $Z_{cs}(4220)$ is definitely a new structure. We would like to call the attention that in the vicinity of the $D_s^* \bar D^*$ threshold there are two points striking out of the LHCb fit, which could be indication of a dynamical structure, which we discuss here.
These points can be observed in the $J/\psi K^+$ distribution of Fig.~3 of Ref.~\cite{LHCb:2021uow} between the two peaks associated to the $Z_{cs}(4000)$ and $Z_{cs}(4220)$ states. These two latter states are assumed to have $J^P = 1^+$ and $1^-$ respectively, so the structure that we obtain with $J^P = 2^+$ cannot be associated to any of these states and is a genuine new structure.

\begin{figure}[h]
 \centering
 \includegraphics[scale=0.5]{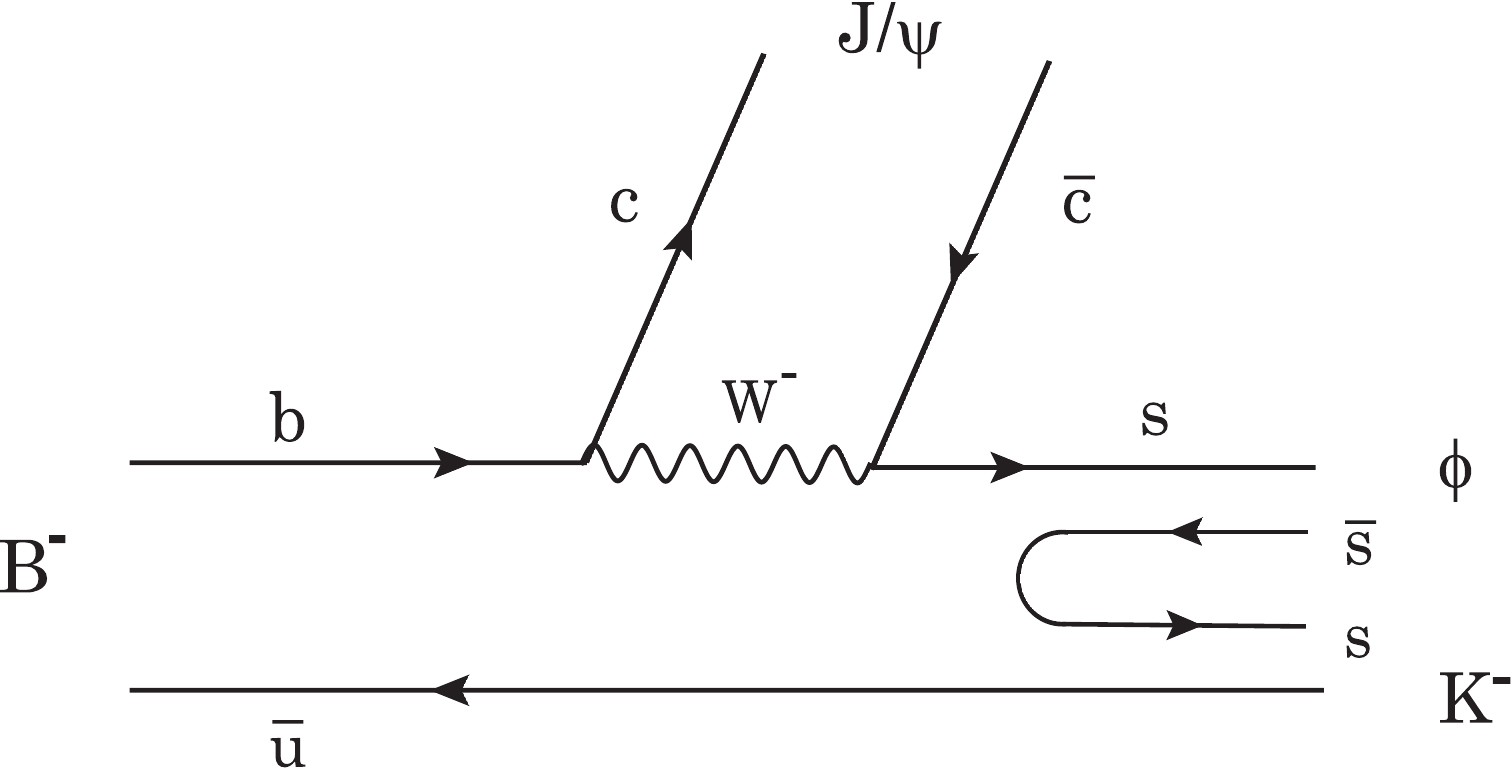}
 \caption{A mechanism for $B^- \to J/\psi \phi K^-$ decay based on internal emission.}
\label{fig:fig5}
\end{figure}

Let us look at how the $B^+ \to J/\psi \phi K^+$ decay proceeds at the microscopical level. We look at the charge conjugate reaction to work with $b$ quarks. In Fig.~\ref{fig:fig5}, we have a mechanism for this decay at the quark level.

\begin{figure}[h]
 \centering
 \includegraphics[scale=0.55]{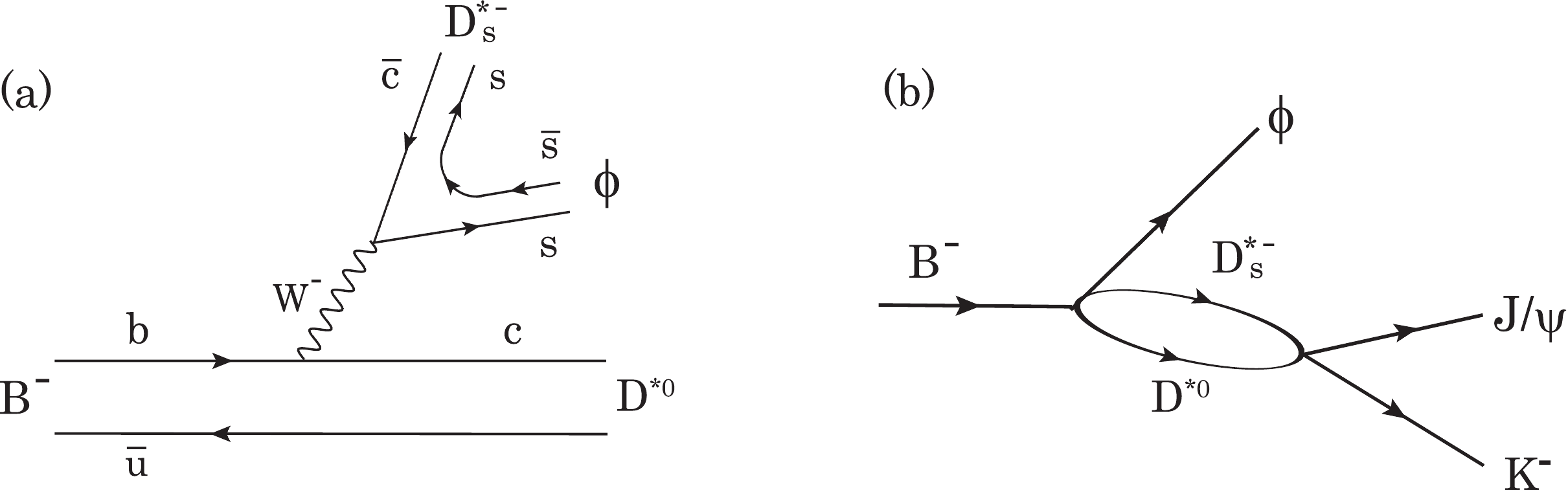}
 \caption{(a) Mechanism for $B^- \to \phi D^{*-}_s D^{*0}$; (b) Rescattering mechanism leading to $\phi J/\psi K$.}
\label{fig:fig6}
\end{figure}

However, we have investigated the $D^*_s \bar D^*$ structures and their decay to $J/\psi K$. Thus, we can also have the mechanism of Fig.~\ref{fig:fig6} which involves external emission and is in principle favored.

\begin{figure}[h]
 \centering
 \includegraphics[scale=0.6]{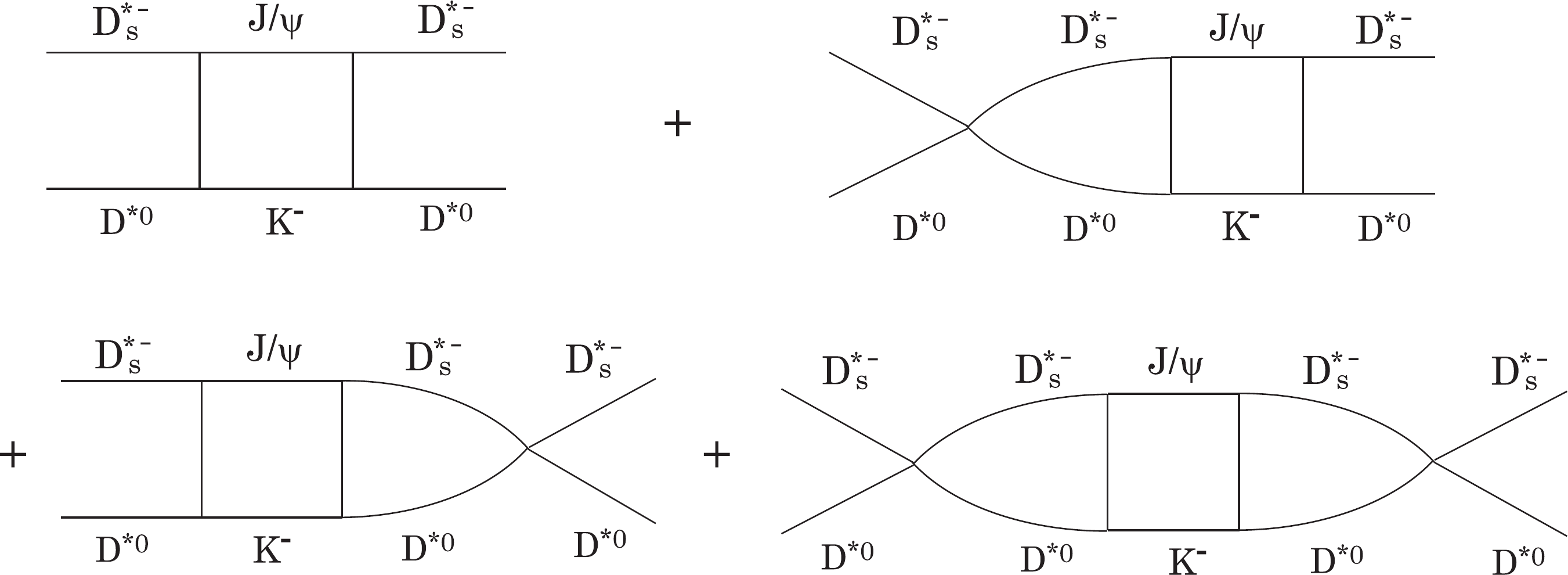}
 \caption{ Diagrams for the transition $D^{*-}_s D^{*0} \to J/\psi K^- \to D^{*-}_s D^{*0}$. Intermediate $K^- J/\psi$ states are implicitly assumed.
}
\label{fig:fig7}
\end{figure}

The mechanism of Fig.~\ref{fig:fig6} produces $\phi D^{*-}_s D^{*0}$, but upon rescattering of $D^{*-}_s D^{*0} \to J/\psi K^-$, as shown in Fig.~\ref{fig:fig6}~(b), we can have $J/\psi \phi K^-$ at the end. This mechanism would reveal any structure tied to a possible $D^{*-}_s \bar D^{*0}$ molecular state. Indeed, the $B^-$ decay amplitude will have two structures, which ignoring the spin dependence will read as,
\begin{eqnarray}
 t^{(1)} &=& A', \\
 t^{(2)} &=& B' \, G_{D^{*-}_s D^{*0}} \, t_{D^{*-}_s D^{*0}, J/\psi K^-} \,,
\end{eqnarray}
and we will assume that they do not interfere. Actually for the only relevant case of $J=2$ for $D^{*}_s D^{*0}$, the structure of $t^{(2)}$ is quite different form the $s$-wave $\vec \epsilon \,(J/\psi) \vec \epsilon \,(\phi)$ of $t^{(1)}$. Then we will have a structure at the end for $|t|^2$ summed over spin polarization of the vectors as,
\begin{equation}
 |t|^2 = |A|^2 + |B|^2 | G_{D^{*-}_s D^{*0}}|^2  | t_{D^{*-}_s D^{*0}, J/\psi K^-}|^2,
\label{eq:t2}
\end{equation}
where $t_{D^{*-}_s D^{*0}, J/\psi K}$ is a transition matrix from the state $D^{*-}_s D^{*0}$ to $J/\psi K^-$. We can obtain this transition matrix from the previous evaluation of Im$V^{'(a)}$ of Eq.~(\ref{eq:ImVpa}). This latter magnitude was evaluated from the diagram of Fig.~\ref{fig:fig4} and was included as a source of potential, see Eq.~(\ref{eq:V_all}), in the evaluation of the Bethe-Sapeter equation. This equation will generate the diagrams of Fig.~\ref{fig:fig7}
which can be summed up as \footnote{We simplify the formalism and ignore the $J/\psi K^*$ channel. The purpose is to show that $t_{D^{*-}_s D^{*0}, J/\psi K^-}$ is proportional to $t_{D^{*-}_s D^{*0}, D^{*-}_s D^{*0}}$.}
\begin{equation}
 ( 1 + G_{D^{*-}_s D^{*0}} \, t_{D^{*-}_s D^{*0}, D^{*-}_s D^{*0}} )   \, V_{D^{*-}_s D^{*0}, J/\psi K^-} \, G_{J/\psi K^-} \, V_{D^{*-}_s D^{*0}, J/\psi K^-} \, 
 ( 1 + G_{D^{*-}_s D^{*0}} \, t_{D^{*-}_s D^{*0}, D^{*-}_s D^{*0}} )
\end{equation}
which upon the use of the Bethe-Salpeter equation,
\begin{equation}
 t_{D^{*-}_s D^{*0}, D^{*-}_s D^{*0}} = V_{D^{*-}_s D^{*0}, D^{*-}_s D^{*0}}
+ V_{D^{*-}_s D^{*0}, D^{*-}_s D^{*0}} \,  G_{D^{*-}_s D^{*0}} \, t_{D^{*-}_s D^{*0}, D^{*-}_s D^{*0}}
\end{equation}
can be rewritten as
\begin{equation}
 \frac{t_{D^{*-}_s D^{*0}, D^{*-}_s D^{*0}} }{V_{D^{*-}_s D^{*0}, D^{*-}_s D^{*0}}}
\, V_{D^{*-}_s D^{*0}, J/\psi K^-} \,  G_{J/\psi K^-} \, V_{D^{*-}_s D^{*0}, D^{*-}_s D^{*0}} \, \frac{t_{D^{*-}_s D^{*0}, D^{*-}_s D^{*0}} }{V_{D^{*-}_s D^{*0}, D^{*-}_s D^{*0}}} \,.
\end{equation}
If we evaluate the imaginary part of this magnitude corresponding to placing $J/\psi K^-$ on shell we get
\begin{equation}
 \left| \frac{t_{D^{*-}_s D^{*0}, D^{*-}_s D^{*0}} }{V_{D^{*-}_s D^{*0}, D^{*-}_s D^{*0}}}
\right|^2 \left( V_{D^{*-}_s D^{*0}, J/\psi K^-} \right)^2 \mathrm{Im} G_{J/\psi K^-}
\end{equation}  
which corresponds to 
\begin{equation}
 \left| t_{D^{*-}_s D^{*0}, J/\psi K^-} \right|^2  \mathrm{Im} G_{J/\psi K^-} \, ,
\end{equation}
with
\begin{equation}
 \mathrm{Im} G_{J/\psi K^-} = - \frac{1}{8\pi} \frac{1}{M_\mathrm{inv}(J/\psi K)} \, q,~~~~~~
 q = \frac{\lambda^{1/2}(M^2_\mathrm{inv}(J/\psi K), M^2_{J/\psi}, M^2_K ) }{2 M_\mathrm{inv}(J/\psi K)}
\end{equation}
Since
\begin{equation}
 \left( V_{D^{*-}_s D^{*0}, J/\psi K^-} \right)^2 \mathrm{Im} G_{J/\psi K^-} = \mathrm{Im} V^{'(a)} + \mathrm{Im} V^{'(b)}
\end{equation}
Eq.~(\ref{eq:t2}) will become
\begin{equation}
 |t|^2 = |A|^2 + |B|^2 | G_{D^{*-}_s D^{*0}}|^2 
 \left| \frac{t_{D^{*-}_s D^{*0}, D^{*-}_s D^{*0}} }{V_{D^{*-}_s D^{*0}, D^{*-}_s D^{*0}}}
\right|^2  
\cdot (-) \frac{8\pi M_\mathrm{inv}(J/\psi K)}{q} (\mathrm{Im} V^{'(a)} + \mathrm{Im} V^{'(b)})
\end{equation}

The mass distribution for $B^- \to J/\psi \phi K^-$ is then given by 
\begin{equation}
 \frac{d \Gamma}{d M_{\rm inv}(J/\psi K)} = \frac{1}{(2\pi)^3} \frac{1}{4 M^2_B} \, p_\phi \, \tilde p_K  \, |t|^2
\label{eq:dGam}
\end{equation}
with
\begin{equation}
 p_\phi = \frac{\lambda^{1/2}(M^2_B,  M^2_\phi, M^2_\mathrm{inv}(J/\psi K)) }{2 M_B},~~~~~~
\tilde p_K =  \frac{\lambda^{1/2}(M^2_\mathrm{inv}(J/\psi K), M^2_{J/\psi}, M^2_K ) }{2 M_\mathrm{inv}(J/\psi K)}
\end{equation}

\section{Results}

With the potential obtained in the former section, we solve now the Bethe-Salpeter equation in coupled channels,
\begin{equation}
 T= [1 -VG]^{-1} V,
\end{equation}
where $G$ is the diagonal meson baryon loop function
\begin{eqnarray}
G = \left(
\begin{array}{cc}
G_{D^*_s D^{*0}} & 0 \\
0 & G_{J/\psi K^*} \\
\end{array}
\right),
\end{eqnarray}
for which we take the formula with cut off method,
\begin{equation}
G_l=\int\frac{d^3q}{(2\pi)^3}\,\frac{\omega_1+\omega_2}{2\omega_1\omega_2}\,\frac{1}{(P^0)^2-(\omega_1+\omega_2)^2+i\epsilon}\ ,
\label{eq:Gfunc}
\end{equation}
with $\omega_1=\sqrt{m_1^2+\vec{q}^{\ 2}}$, $\omega_2=\sqrt{m_2^2+\vec{q}^{\ 2}}$ ($m_1$ and $m_2$ are the vector masses of the $l$ channel). 

As for the cut-off parameters of the $G$ function in Eq.~(\ref{eq:Gfunc}), we use different values of $q_{\rm max}$ and $q'_{\rm max}$ for the different channels. For $q_{\rm max}$ we use values between 450~MeV and 650~MeV.
Values around $q_{\rm max}= 700-850$~MeV were used in Ref.~\cite{ikeno} to get an enhancement in the $\bar D_s D^* + \bar D^*_s D$ mass distribution close to threshold which was proposed as an explanation of the $Z_{cs}(3985)$. On the other hand, values around 420--450~MeV are used in Ref.~\cite{feijoo} to explain the $T_{cc}$ state as a molecule of $D^* D$.
For the $J/\psi K^*$ channel we take a larger value of $q'_{\rm max}$ to avoid having the on shell $J/\psi K^*$ momentum at energies close to the $D^*_s \bar D^*$ threshold bigger than the cut off. The results are shown for different values of $q'_{\rm max}$ ranging from 700--900~MeV.

As shown in Eqs.~(\ref{eq:Vc_J0}) to (\ref{eq:Vex_J1}), the potential for different spins, $J=0,1,2$ are different. We see that $J=1$ is the most unlikely case to develop a bound state because the contact term is repulsive. The attraction from $V^{\rm (ex)}$, coming from $J/\psi$ exchange is very small and there is no connection between $D^*_s \bar D^*$ and $J/\psi K^*$. The case of $J=0$ is more favorable because now there is a coupled channel effect from a non vanishing $D^*_s \bar D^* \to J/\psi K^*$ transition. Yet, the diagonal interaction is very weak and the contact term is still repulsive. On the other hand, the $J=2$ case is the most favorable, since the contact term is attractive and one also has the $D^*_s \bar D^* \to J/\psi K^*$ transition. We should note that all the interaction terms are subleading in the heavy quark counting and do not follow the heavy quark spin symmetry rules. Indeed, both $J/\psi$ or $D^*_s$, $D^*$ exchange are subleading with respect to light vector exchange. In this latter case the heavy quarks are spectators in the vector exchange process and hence the matrix elements are independent on the heavy quarks. Thus, the light vector exchange terms automatically fulfill heavy quark spin symmetry, but not the other terms where one exchanges some heavy quarks. The contact term is also subleading since it is not proportional to the external energies of the mesons, unlike the vector exchange. Because all the terms of the interaction are small, it is unlikely that one can form a bound state of the system, but like the case of the $Z_{cs}(3985)$ we could also have cusp effects around the $D^*_s \bar D^*$ threshold.

\begin{figure}[tb]
\centering
\includegraphics[scale=0.55]{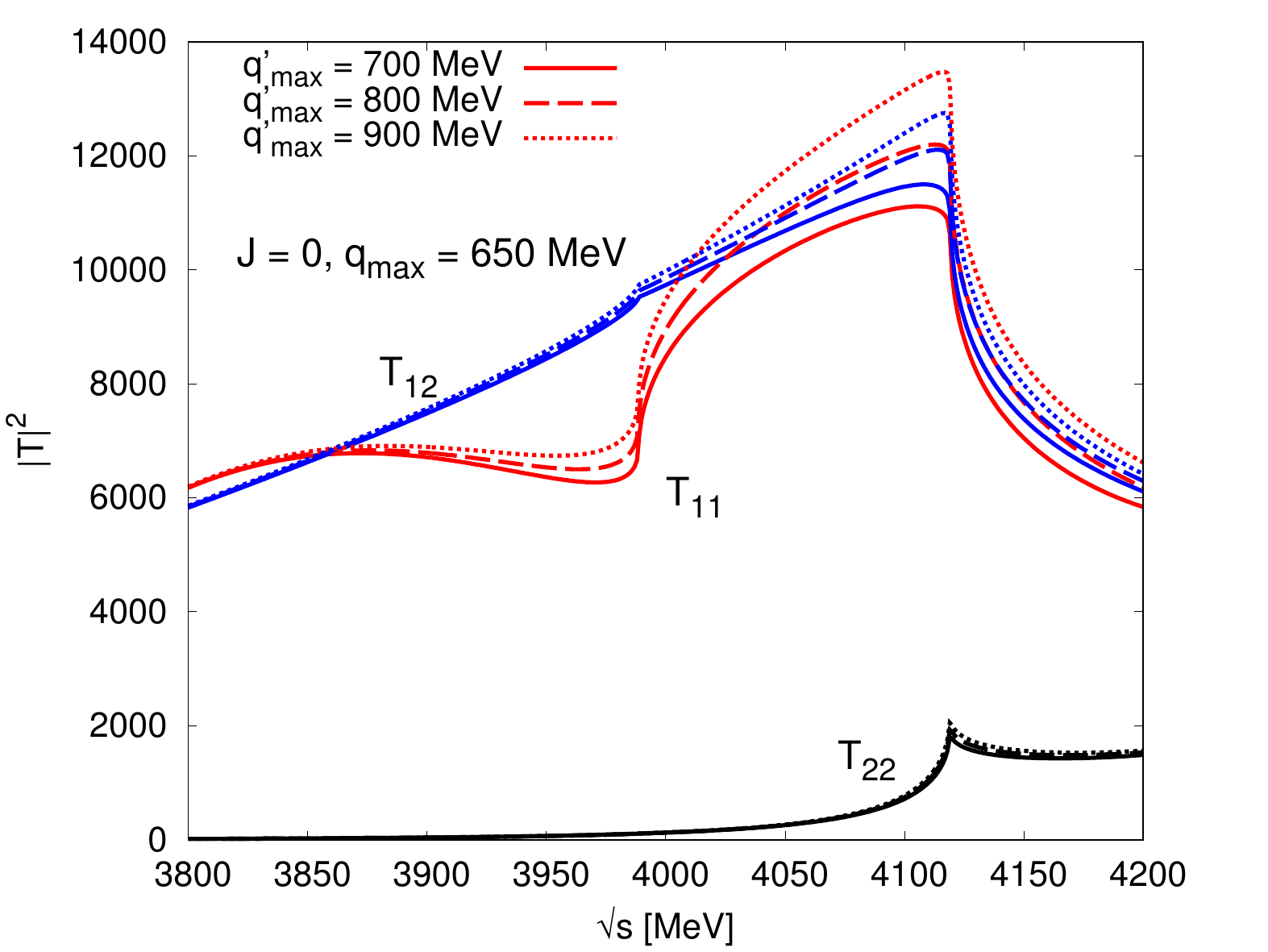}
\includegraphics[scale=0.55]{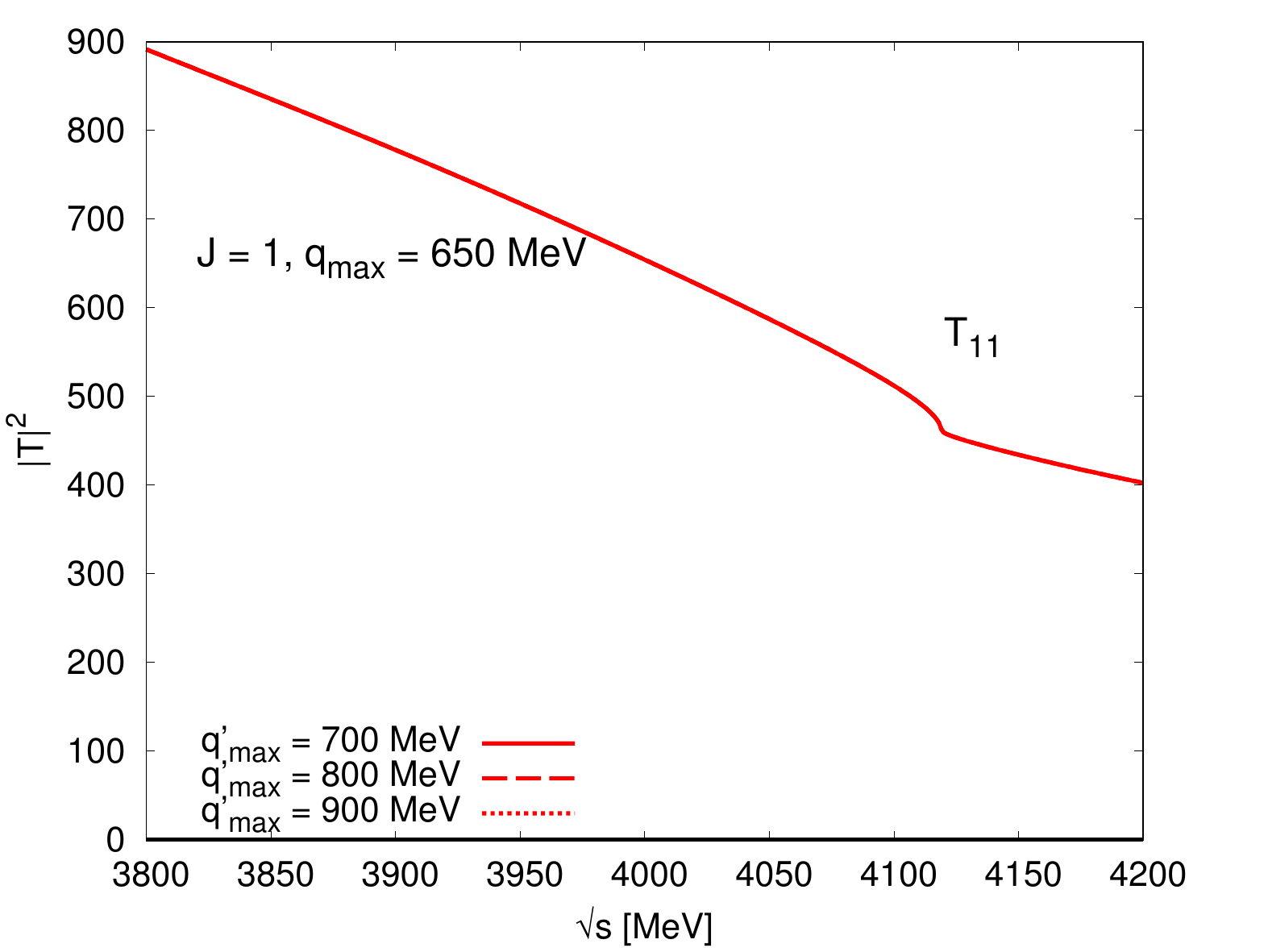}
\includegraphics[scale=0.55]{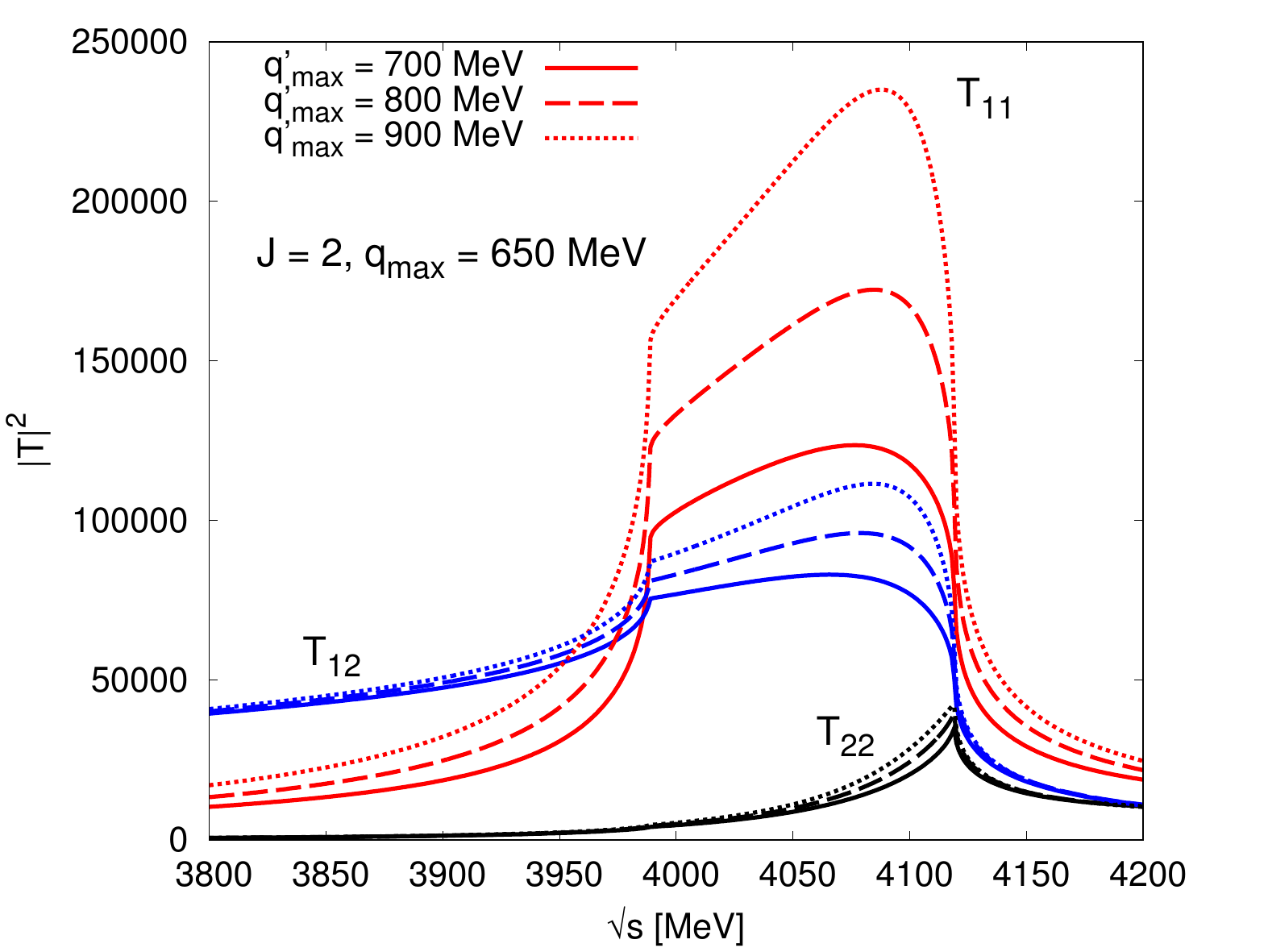}
 \caption{ $|T|^2$ for the each channel are shown for the spin $J$ and the different $q'_{\rm max}$ value of the $J/\psi K^*$ channel. The value of $q_{\rm max} = 650$~MeV is fixed for the $D^{*}_s \bar D^*$ channel. 
}
\label{fig:Tmax_qmax650}
\end{figure}

\begin{figure}[tb]
\centering
\includegraphics[scale=0.55]{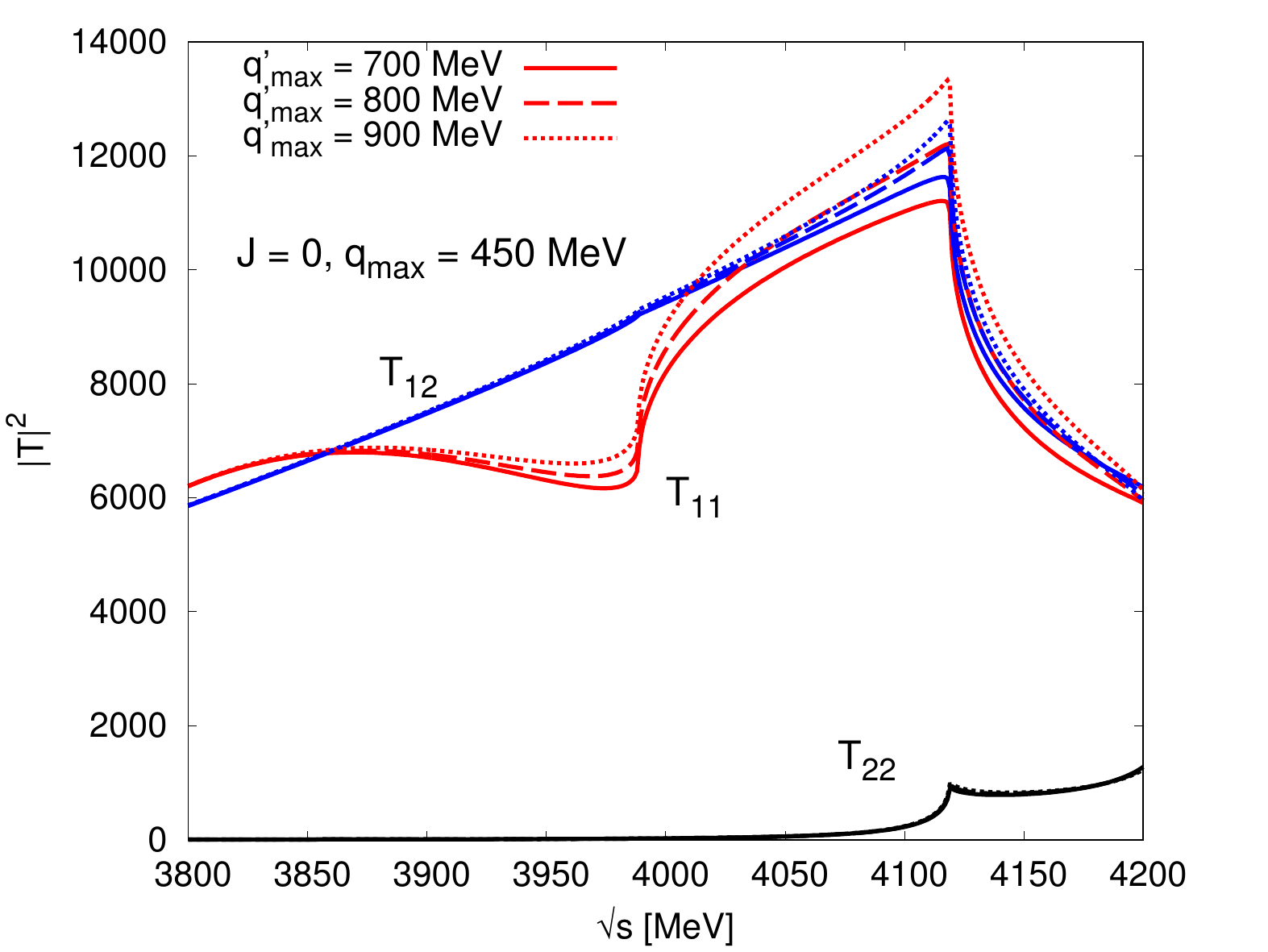}
\includegraphics[scale=0.55]{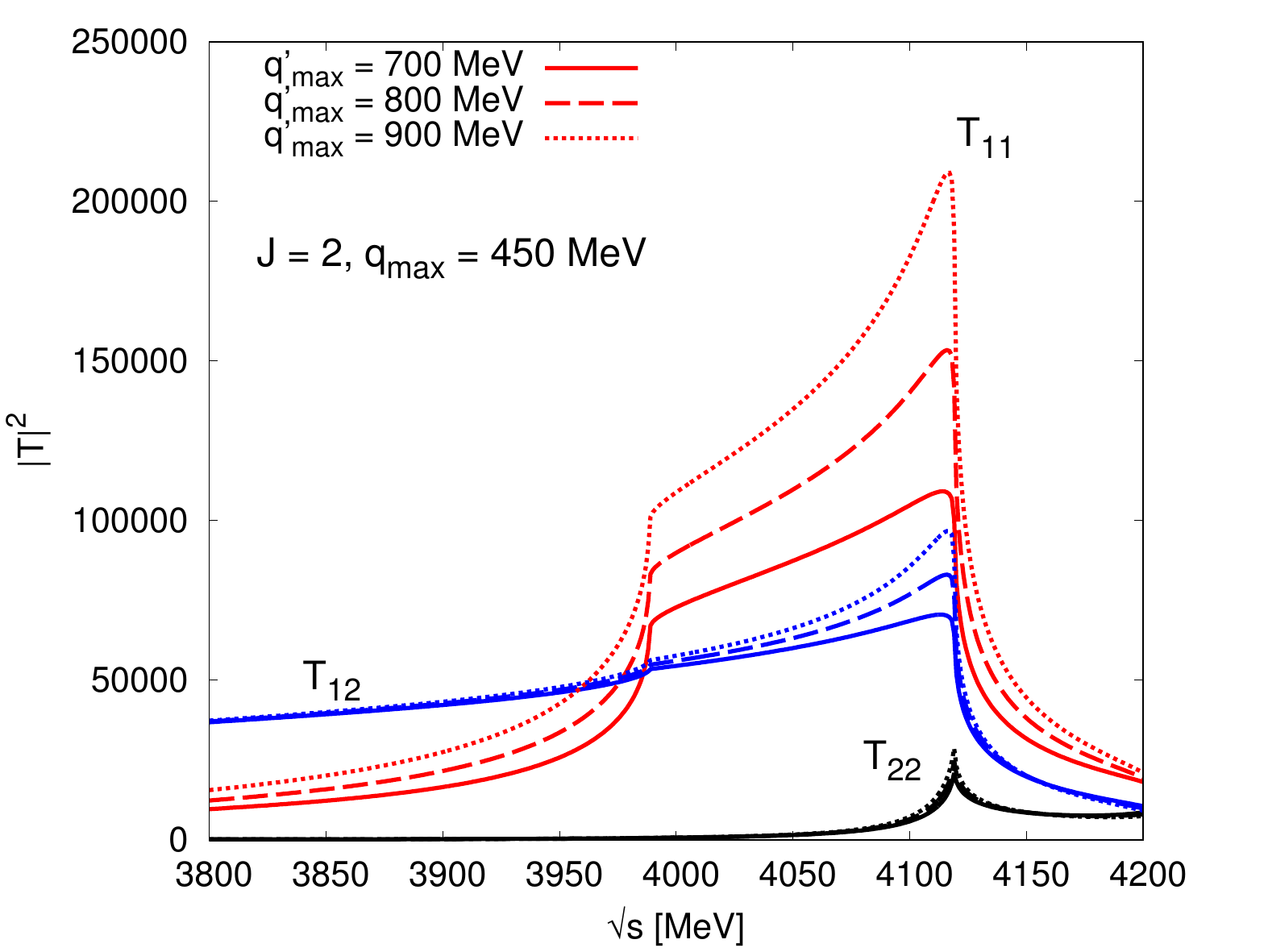}
\caption{ $|T|^2$ for the each channel are shown for the spin $J$ and the different $q'_{\rm max}$ value of the $J/\psi K^*$ channel. The value of $q_{\rm max} = 450$~MeV is fixed for the $D^{*}_s \bar D^*$ channel. 
}
\label{fig:Tmax_qmax450}
\end{figure}

We can see the repercussions of the former discussion in the values of $|T|^2$ which we show in Fig.~\ref{fig:Tmax_qmax650}. Indeed, we can see that the strength of $|T|^2$ for $J=1$ in the diagonal $D^*_s \bar D^* \to D^*_s \bar D^*$ transition is very weak and only a tiny cusp is seen in the $D^*_s \bar D^*$ threshold. 
On the other hand, a cusp like structure is seen for $|T_{11}|^2$ for $J=0$ both at the $J/\psi K^*$ and $D^*_s \bar D^*$ thresholds, particularly in the second channel. The same occurs for $J=2$, but here the strength of $|T_{11}|^2$ is a factor of 15 times larger than for $J=0$. Because of that, we should associate any structure observed at the $D^*_s \bar D^*$ threshold to $J=2$. In Fig.~\ref{fig:Tmax_qmax450} we show again $|T_{11}|^2$ for $J=0$ and $J=2$ for $q_{\rm max}=450$~MeV instead of 650~MeV used in Fig.~\ref{fig:Tmax_qmax650}. We observe that the cusp structure around the two thresholds is very similar and the strength of the magnitude has not changed much. Particularly visible is the cusp structure around the $D^*_s \bar D^*$ threshold (4119~MeV). This structure is typical of a barely ``missed'' bound state, or virtual state.
The reason why in the case of $J=1$ there is only one curve independent on $q'_{\rm max}$ is that, as one can see in Eqs.~(\ref{eq:Vc_J1}) and (\ref{eq:Vex_J1}),
the matrix elements including the $J/\psi K^+$ state are all zero and hence the loops including $q'_{\rm max}$ do not appear in the scheme.

\begin{figure}[h]
\centering
\includegraphics[scale=0.55]{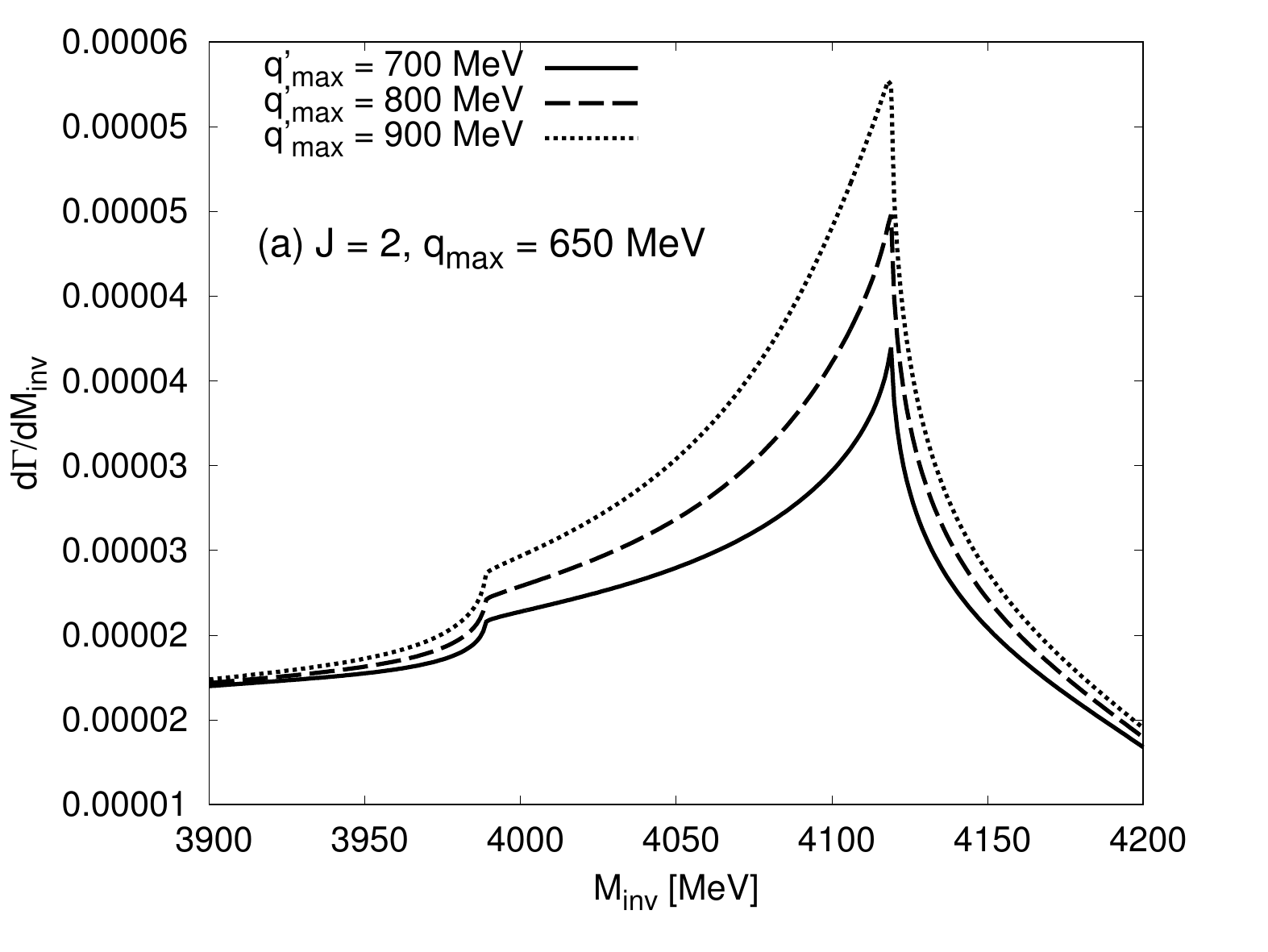}
\includegraphics[scale=0.55]{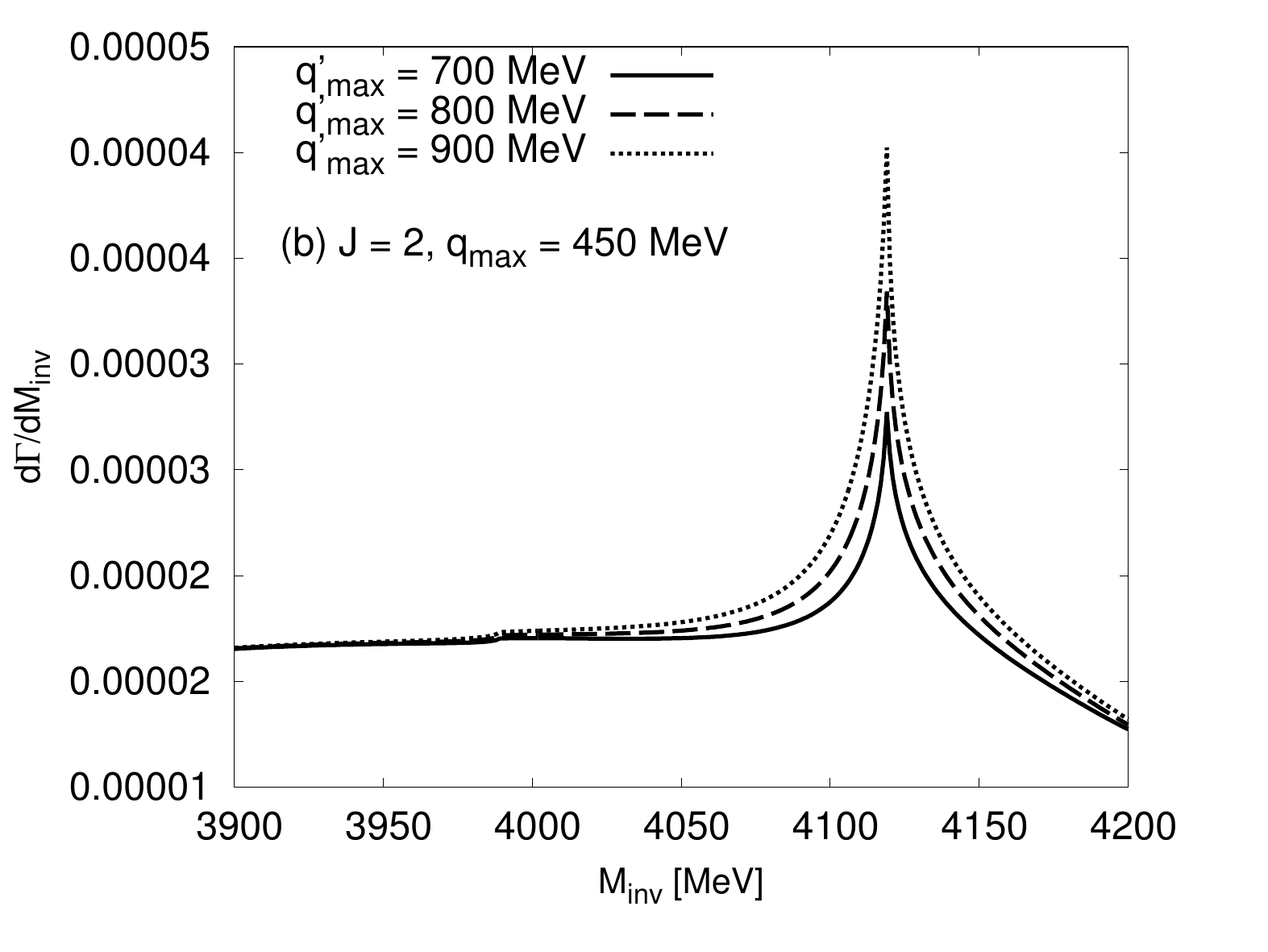}
\caption{
The mass distribution for $B^- \to J/\psi \phi K^-$ is show as a function of $M_\mathrm{inv}(J/\psi K)$. 
(a) $q_{\rm max} = 650$~MeV and (b) $q_{\rm max} = 450$~MeV are fixed for the $D^{*}_s \bar D^*$ channel and the different values of $q'_{\rm max}$ for the $J/\psi K^*$ channel are used as shown in the figure. The result corresponds to the case of $J=2$ for $D^*_s D^{*0}$. We use $A=1$ and $B=3A$, as approximately dictated by color counting in internal and external emissions.
}
\label{fig:dGam_J2}
\end{figure}

\begin{figure}[h]
\centering
\includegraphics[scale=0.55]{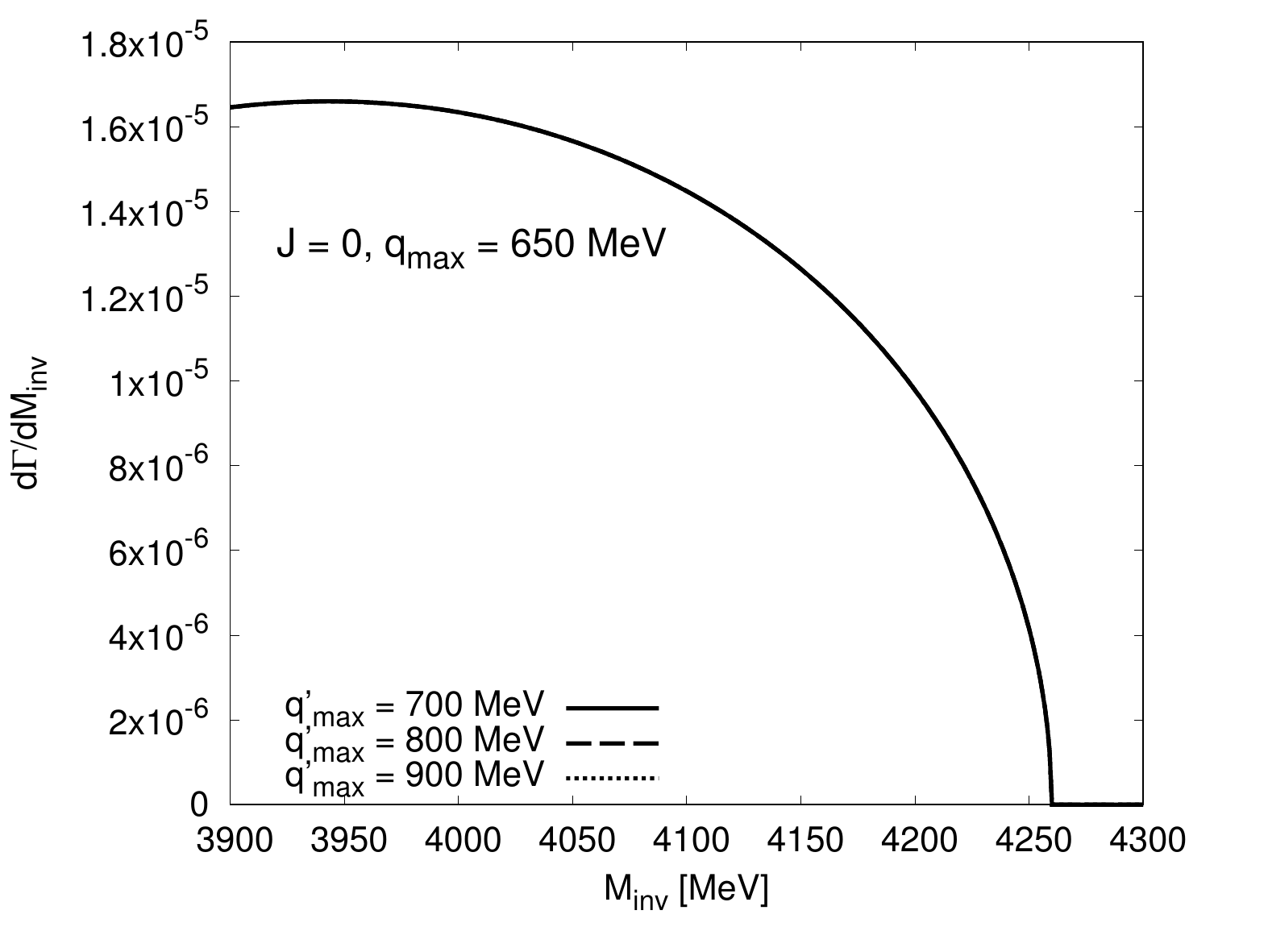}
\includegraphics[scale=0.55]{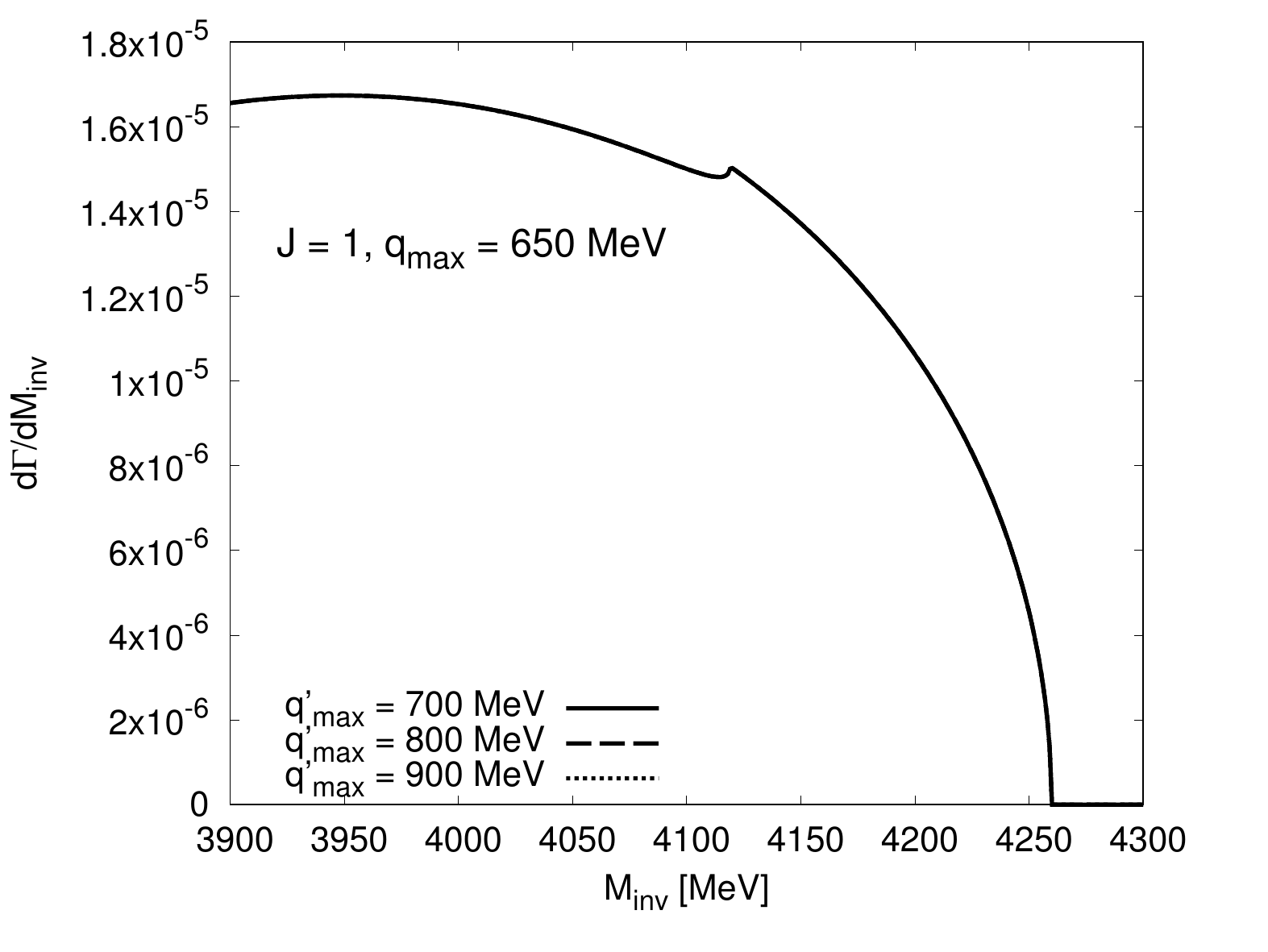}
\caption{
Same as Fig.~\ref{fig:dGam_J2} except for the cases of $J=0$ and $J=1$ for $D^*_s D^{*0}$.
}
\label{fig:dGam_J01}
\end{figure}

Finally we would like to show the results for the $J/\psi K^-$ distribution in the $B^- \to J/\psi \phi K^-$ decay. In Fig.~\ref{fig:dGam_J2}, we show the mass distribution for the case of $J=2$ based on Eq.~(\ref{eq:dGam}). We see a pronounced sharp peak around the $D^*_s \bar D^*$ threshold on top of a background created by the tree level ($|A|^2$ term of Eq.~(\ref{eq:t2})) which should be visible in an experiment. In Fig.~\ref{fig:dGam_J2}~(a) the results are shown for $q_{\rm max}=650$~MeV and different values of $q'_{\rm max}$, while in Fig.~\ref{fig:dGam_J2}~(b) we show results for $q_{\rm max}=450$~MeV. The features are qualitatively similar.
In Fig.~\ref{fig:dGam_J01}, we show the same mass spectrum for $J=0$ and $J=1$. 
As we can see, there is no signal for $J=0$, since $W'_s=0$ for $J=0$ (see Eq.~(\ref{eq:Wps})) and the cusp effect at the $D^*_s \bar D^{*}$ threshold for $J=1$ is negligible, indicating that any possible experimental signal there should be attributed to a $J = 2$ state.
For the same reasons as in Fig.~\ref{fig:Tmax_qmax650}, commented before, the results for $J=1$ do not depend on $q'_{\rm max}$.

It is interesting to mention that the data of $B^+ \to J/\psi \phi K^+$ in Ref.~\cite{LHCb:2021uow} (see Fig.~3 of that reference) have two points sticking out of the bulk of the data around the $D^*_s \bar D^*$ threshold. This could also be a statistical fluctuation, but the observation becomes more relevant when one realizes, as remarked in Ref.~\cite{xucao}, that the $\bar B^0_s \to J/\psi K^- K^+$ decay~\cite{jpsikk} shows also a sharp peak in the spectrum at precisely the $D^*_s \bar D^{*0}$ threshold, based on one point clearly deviating much more than the errors and fluctuations from the bulk of the data
(see Fig.~16, left of Ref.~\cite{jpsikk} where a peak is seen in the $J/\psi K^+$ invariant mass distribution around 4120~MeV).
The signals in the two independent reactions at the same invariant mass are unlikely to be accidental and deserve more attention.

\section{Conclusions}
We have studied the $D_s^{*+} \bar D^{*0}$ system in connection with $J/\psi K^{*+}$ with coupled channels, using the extension of the local hidden gauge approach to obtain the interaction. The Bethe-Salpeter equation is used to generate the scattering matrix of the channels and we observe that, with natural values of the cut off parameters used to regularize the loops, we do not obtain bound states for the system. We observe that in the $J^P=1^+$ channel the repulsion dominates and we see no structure in the amplitudes. The $J^P = 0^+$ channel exhibits some cusp structure around the two channels thresholds, which is more visible in the $J^P =2^+$ channel, that has a strength in $|T|^2$ about 15 times larger than for $J^P = 0^+$ as a consequence of a stronger attraction in the interaction. Even then one fails to obtain a bound state of the system, a situation which is very similar to the one of the $a_0(980)$ resonance, very clearly seen in experiments with a cusp structure~\cite{BESIII:2016tqo}, as well as in the theoretical description in the chiral unitary approach~\cite{lianga0}. Yet, as in the case of the $a_0(980)$, accepted commonly as a resonance, a strong cusp around the threshold of a hadron-hadron state is indicative of a particular dynamics of this hadron-hadron system, in this case an attractive interaction, that fails for short to produce a bound state and gives instead rise to a virtual state~\cite{guozou}.

We showed that the structure seen in $|T|^2$ of the $J^P = 2^+$ channel had repercussion in the theoretical prediction of the $J/\psi K^+$ spectrum in the $B^+ \to J/\psi \phi K^+$ reaction, in terms of a sharp peak in the $J/\psi K^+$ invariant mass distribution around the $D_s^* \bar D^{*0}$ threshold. Interestingly, a possible peak structure at that energy was visible in the experiment~\cite{LHCb:2021uow}, and also a similar peak was visible in the $J/\psi K^+$ invariant mass distribution at the same energy in the $\bar B^0_s \to J/\psi K^- K^+$ reaction~\cite{jpsikk}. Such a coincidence can hardly be accidental and should serve as a motivation to measure with precision this energy region in future experiments with the conviction that a cusp structure bears as much information on hadron dynamics as actual resonances.


\section*{ACKNOWLEDGMENT}
The work of N. I. was partly supported by JSPS KAKENHI Grant Number JP19K14709. 
R. M. acknowledges support from the CIDEGENT program with Ref. CIDEGENT/2019/015 and from the spanish national grants PID2019-106080GB-C21 and PID2020-112777GB-I00. 
This work is partly supported by the Spanish Ministerio de Economia y Competitividad and European FEDER funds under Contracts No. FIS2017-84038-C2-1-P B and by Generalitat Valenciana under contract
PROMETEO/2020/023. This project has received funding from the European Unions Horizon 2020 research and innovation programme under grant agreement No. 824093 for the “STRONG-2020” project.

\end{document}